z
\documentclass[%
reprint,
 amsmath,amssymb,
 aps,
]{revtex4-2}

\usepackage{graphicx}
\usepackage{dcolumn}
\usepackage{bm}
\usepackage{hyperref}
\usepackage{float}
\usepackage{booktabs}

\usepackage{changes}

\begin{document}

\preprint{APS/123-QED}

\title{Solar-wind electron precipitation on weakly magnetized bodies: \\ the planet Mercury}

\author{Federico Lavorenti}
\email{federico.lavorenti@oca.eu}
\affiliation{Laboratoire Lagrange, Observatoire de la Côte d’Azur, UCA, CNRS, Nice, France}
\affiliation{Dipartimento di Fisica ``E. Fermi'', Università di Pisa, Pisa, Italy}

\author{Pierre Henri}
\affiliation{Laboratoire Lagrange, Observatoire de la Côte d’Azur, UCA, CNRS, Nice, France}
\affiliation{LPC2E, CNRS, Univ. d’Orléans, OSUC, CNES, Orléans, France}

\author{Francesco Califano}
\affiliation{Dipartimento di Fisica ``E. Fermi'', Università di Pisa, Pisa, Italy}

\author{Jan Deca}
\affiliation{LASP, University of Colorado Boulder, Boulder, CO, USA}
\affiliation{Institute for Modeling Plasma, Atmospheres and Cosmic Dust, NASA/SSERVI, Silicon Valley, CA, USA}
\affiliation{LATMOS, Université de Versailles à Saint Quentin, Guyancourt, France}

\author{Simon Lindsay}
\affiliation{School of Physics \& Astronomy, University of Leicester, Leicester, UK}

\author{Sae Aizawa}
\affiliation{IRAP, CNRS-CNES-UPS, Toulouse, France}
\affiliation{Dipartimento di Fisica "E. Fermi", Università di Pisa, Pisa, Italy}
\affiliation{Japan Aerospace Exploration Agency, Sagamihara, Kanagawa, Japan}

\author{Johannes Benkhoff}
\affiliation{ESA/ESTEC, Keplerlaan 1, 2200 AG Noordwijk, The Netherlands}

\date{\today}

\begin{abstract}
Mercury is the archetype of a weakly magnetized, airless, telluric body immersed in the solar wind. Due to the lack of any substantial atmosphere, the solar wind directly precipitates on Mercury's surface.
Using a 3D fully-kinetic self-consistent plasma model, we show for the first time that solar-wind electron precipitation drives (i) efficient ionization of multiple species (H, He, O and Mn) in Mercury's neutral exosphere and (ii) emission of X-rays from the planet's surface.
This is the first, independent evidence of X-ray auroras on Mercury using a numerical approach.
\end{abstract}

\maketitle

\begin{table}
    \centering
    \begin{tabular}{ccc}
        \toprule
        Box dimensions                  & $(L_x,L_y,L_z)$   & (-9:+6,$\pm$6,$\pm$6)~$R$ \\
        Number of cells                 & $(N_x,N_y,N_z)$   & (960,768,768)            \\
        Spatial resolution              & $(dx,dy,dz)$      & 0.015~$R$                 \\
        Time resolution                 & $dt$              & 1.4 ms                   \\
        Macro-particles per cell\footnote{Total number of macro-particles, ions plus electrons.} 
                                        & PPC               & 128                       \\
        \midrule
        Solar-wind number flux          & $\mathcal{F}_{_{SW}}$ & $1.2\cdot 10^9$ 1/cm$^{2}$s \\
        Solar-wind energy flux          & $\mathcal{E}_{_{SW}}$ & $2.4\cdot 10^{10}$ eV/cm$^{2}$s      \\
        \midrule
        Ion-to-electron mass ratio      & $m_i/m_e$         & 100                      \\ 
        Light-to-Alfv\'en speed ratio   & $c/V_{A,i}$       & 178                      \\
        Planet-to-gyro radius ratio     & $R/\rho_i$        & 10                       \\
        \bottomrule
    \end{tabular}
    \caption{Common numerical parameters of RunN and RunS with purely northward and southwards IMF, respectively.}
    \label{tab:tab1_pars}
\end{table}

The planet Mercury is our neighbor example of a rocky, weakly magnetized body immersed in the solar-wind plasma.
Mercury's environment presents us the ideal scenario to better understand the physics governing the interaction between rocky bodies (such as planets, moons, asteroids and comets) and the solar wind.
Mercury's complexity is due to the strong coupling between the solar-wind plasma and the planet's magnetosphere, exosphere and surface. The physical processes controlling the electron interactions in the system are a scientific enigma with underlying roots ranging from plasma to solid-state physics~\citep{Milillo2010}. As an example, electron acceleration in the magnetosphere is thought to be at the origin of X-ray aurora-like emission from the surface of Mercury. However, on the one hand, this hypothesis remains to be confirmed; on the other hand, the dependence of such process on solar-wind conditions remains unknown.

The Sun acts as an external energy driver, sustaining the dynamics at Mercury. Solar radiation and particles are the source (via sputtering and desorption) and sink (via radiation pressure and ionization) of the neutral exosphere surrounding Mercury. Moreover, the shape and dynamics of Mercury's magnetic field is strongly altered by the solar wind. Mercury's intrinsic magnetic field generates a scaled-down, Earth-like magnetosphere able to -- partially -- shield the planet's surface from the impinging solar wind. Part of the solar wind enters the magnetosphere, interacts with Mercury's magnetic field and, given the absence of an atmosphere, precipitates down to the planet's surface.
The study of such plasma precipitation is a key point to understand the strong coupling between Mercury's magnetosphere, exosphere and surface, and around weakly magnetized bodies in general.

To date, only two missions (Mariner 10 and MESSENGER) have been devoted to the exploration of Mercury's environment.
Mariner 10 provided the first electron observations showing electron fluxes in the range $\sim$20-600 eV throughout the magnetosphere~\citep{Ogilvie1977}. Sporadic bursts above tens of keV were also detected~\citep[and references therein]{Wurz2001}.
MESSENGER~\citep{Solomon2007} could not measure the core of the electron distribution function, but it provided direct observations of high-energy electrons above 35 keV~\citep{Ho2011,Ho2012} and indirect observations of suprathermal electrons in the range $\sim$1-10 keV~\citep{Lawrence2015,Baker2016,Dewey2018,Ho2016}. MESSENGER also reported the first evidence of electron-induced X-ray emissions from the surface~\citep{Lindsay2016,Lindsay2022}. 
These missions provided a novel -- but still fragmented -- picture of the solar-wind electron interaction with Mercury and weakly magnetized bodies, in general.
In the near future, the joint ESA/JAXA space mission BepiColombo~\citep{Benkhoff2021} will revolutionize our understanding of Mercury's environment thanks to (i) its two-satellite composition and (ii) its instrumental payload with resolution down to electron kinetic scales~\citep{Milillo2020}. BepiColombo will observe the whole electron~\citep{Saito2010,Huovelin2020} and X-ray spectrum~\citep{Bunce2020} with unprecedented resolution. 
However, the complexity of these measurements is such that only through global models, including the electron dynamics, the true potential of these measurements can be unveiled. 
Our work presents such a model to interpret the fragmented picture left by past MESSENGER observations, while paving the way for the future BepiColombo ones.

In the past, global numerical models of Mercury's interaction with the solar wind have been focused on the ion dynamics~\citep{Kallio2003,Travnicek2010,Richer2012,Fatemi2020}. Such models included self-consistently the ion kinetic physics, but neglected the kinetic physics of electrons (treated as a massless neutralizing fluid). Those models neglect electron acceleration processes and so their feedback on the magnetosphere and surface.
Previous studies that did model electron trajectories, then, prescribed constant electromagnetic fields (i.e. a test-particle approach~\citep{Schriver2011b,Walsh2013}).
Such a method completely neglects the kinetic electron dynamics and their feedback on the large-scale global evolution.
\citet{Schriver2011b} provided a first estimate of electron precipitation maps at Mercury, but their results did not address the energy distribution of electrons at the surface (due to the small statistical sample of test electrons).
To overcome these limitations, in this work we study the precipitation of electrons on Mercury-like bodies using a global fully-kinetic model. Our model includes the electron dynamics self-consistently from the large, planetary scale down to the electron gyro-radius.

We assess electron precipitation at the surface of Mercury under northward and southward solar-wind conditions. Our numerical results are then used to compute (i) the electron impact ionization rates in Mercury's low altitude exosphere, and (ii) the X-ray photon emission profiles from Mercury's surface. This novel, self-consistent approach provides (i) the first estimate of the efficiency of electron impact ionization -- a process usually neglected in exosphere models -- for multiple exospheric species, and (ii) the first estimate of X-ray luminosity of a rocky, weakly magnetized body driven by solar-wind electron precipitation.
\begin{figure}
    \centering
    \includegraphics[width=\linewidth]{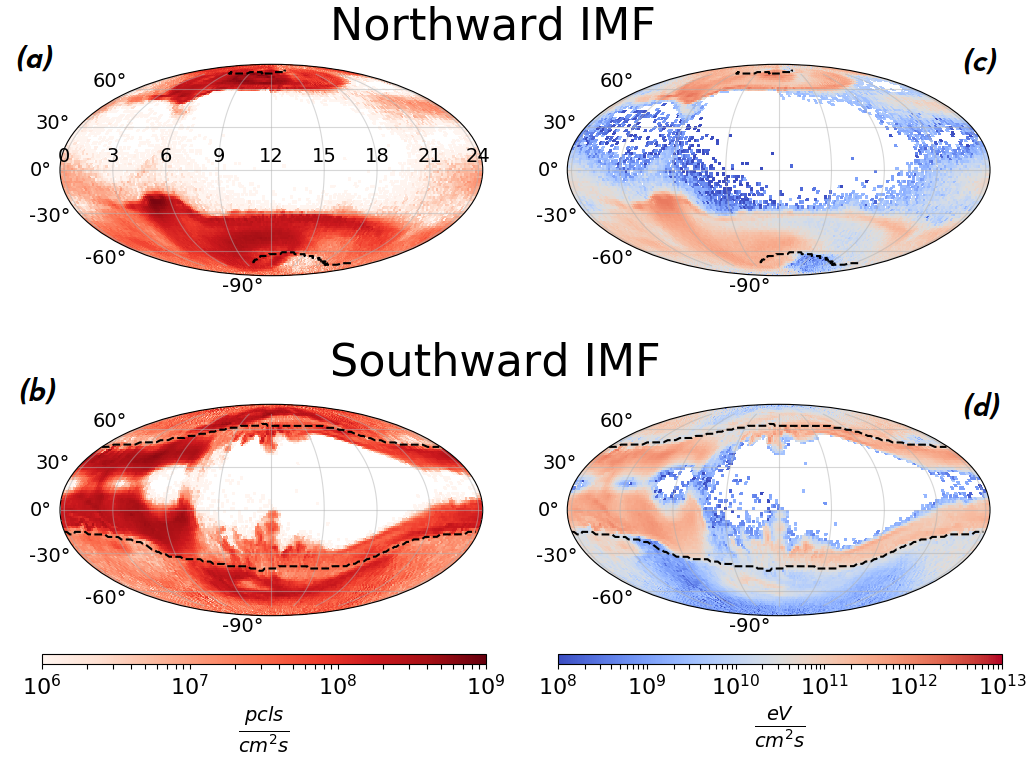}
    \caption{Electron precipitation maps from our fully-kinetic simulations using a purely northward (top panels) or southward (bottom panels) IMF. Panels a-b: number fluxes in units of electrons/cm$^2$s. Panels c-d: energy fluxes in units of eV/cm$^2$s. 
    All maps use a Mollweide projection of the surface of Mercury. 
    Vertical axis corresponds to geographical latitude. Horizontal axis corresponds to local time (LT, as indicated in panel a), where LT 12 is the subsolar longitude. 
    Dashed black lines indicate the boundary between open and closed magnetic field lines.}
    \label{fig:maps_electrons}
\end{figure}
\begin{figure*}
    \centering
    \includegraphics[width=0.63\linewidth]{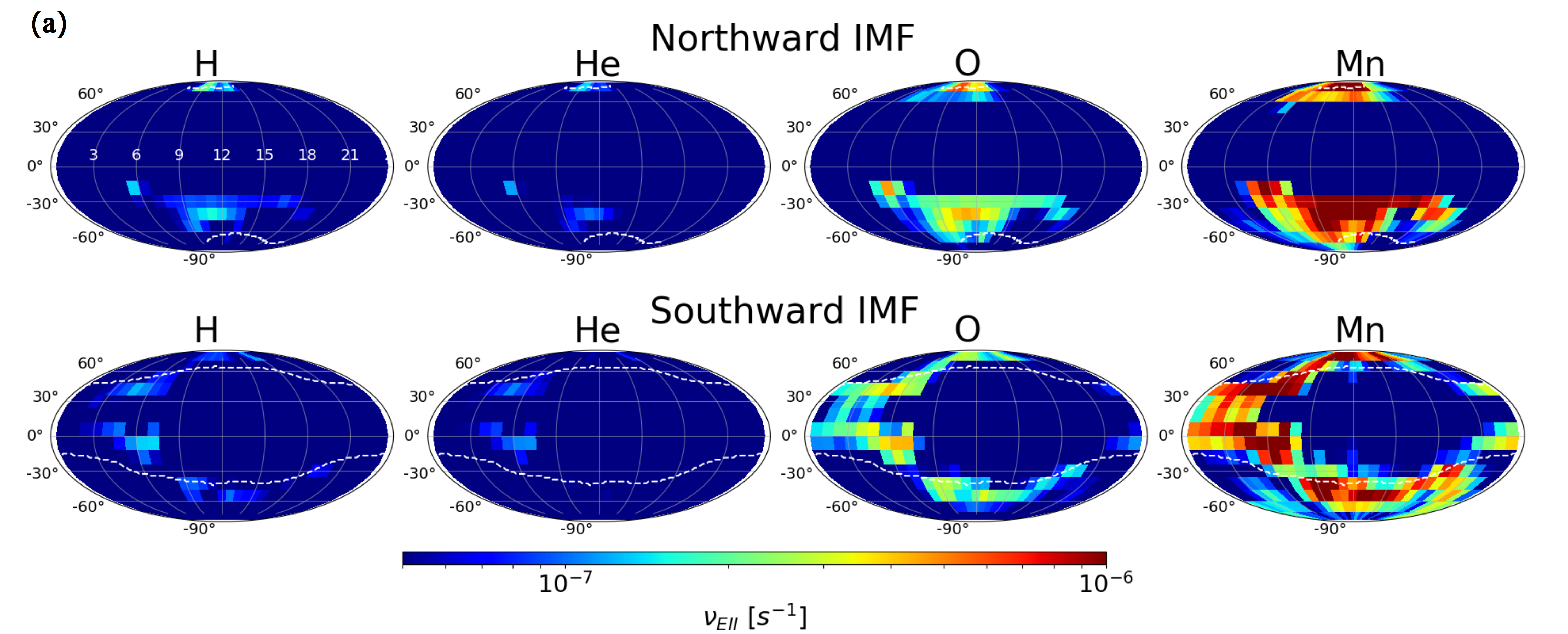}
    \includegraphics[width=0.36\linewidth]{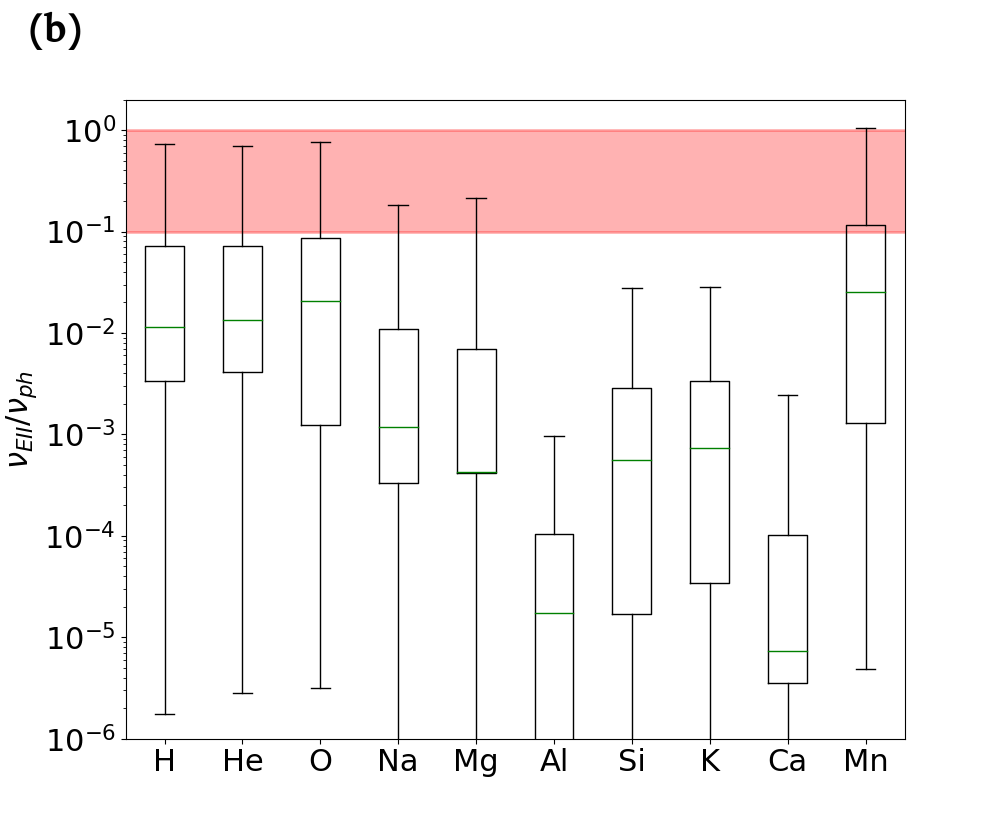}
    \caption{Electron impact ionization (EII) rates computed in our two simulations. 
    Panel a: surface maps of EII rates for selected exospheric species (H, He, O and Mn). Maps in the same format as Fig~\ref{fig:maps_electrons}. 
    Panel b: box-plot of the distribution of EII rates for all exospheric species considered. EII rates are normalized to photoionization rates.
    Horizontal green lines shows the median, limits of the boxes are the first and third quartile, the whiskers are the maxima of the data. 
    In panel b, EII rates obtained from the two runs (RunN and RunS) are merged. 
    The red box in panel b highlights the region of non-negligible EII (i.e. where the EII-to-photoionization frequency ratio is between 0.1 and 1).}
    \label{fig:EII_maps}
\end{figure*}

We use the semi-implicit, particle-in-cell code \texttt{iPIC3D} that solves the Vlasov-Maxwell system of equations in a three-dimensional Cartesian box by discretizing the ion and electron distribution function~\citep{Markidis2010}. The simulation setup includes (i) a uniform solar-wind plasma injected from the sunward side of the box and (ii) a scaled-down model of the planet Mercury with radius $R$ = 230 km (radius reduced by around a factor 10 from its real value) and magnetic dipole moment 200 nT$/R^3$. The dipole field is shifted northward by 0.2~$R$ in agreement with the MESSENGER magnetic field observations~\citep{Anderson2012}. 
A scaled-down setup enables us to run a global fully-kinetic simulation on present state-of-the-art computing facilities. Scaling-down the planet, but keeping the good ordering of physical spatial and temporal scales, preserves the global magnetosphere structure and dynamics~\citep{Lavorenti2022a}.
At the box boundaries, the outermost cells are populated with solar-wind plasma and the electromagnetic fields are linearly smoothed to their solar-wind values. Inside the planet, macro-particles are removed using a charge-balanced scheme~\citep{Lavorenti2022a}.
We initialize the simulations with a solar-wind density $n_{_{SW}}$ = 30 cm$^{-3}$, speed $V_{_{SW}}$ = 400 km/s in the sun-planet direction, a magnetic field amplitude $B_{_{SW}}$ = 20 nT and temperature $T_{i,_{SW}}$ = $T_{e,_{SW}}$ = 21.5 eV. Two different simulation setups, with purely northward (RunN) or southward (RunS) interplanetary magnetic field (IMF) are used.
These two limit cases target the role of magnetic reconnection in the magnetosphere (inhibited in RunN and enhanced in RunS).
This simulation setup has been validated in~\citet{Lavorenti2022a} studying the large-scale structure of Mercury's magnetosphere. The numerical parameters employed in our two runs are reported in Tab.~\ref{tab:tab1_pars}.


In both simulations, the global system reaches a quasi-steady state after a time $T\approx$ 10 s. Starting from this time $T$, we integrate the plasma precipitation on the planet's surface for a time interval $\Delta t$ = 50 ms, corresponding to about two electron gyro-periods.
From the precipitated plasma, we compute the electron precipitation maps, as shown in Fig.~\ref{fig:maps_electrons}~\footnote{Solar-wind ion precipitation maps in the same format as Fig.~\ref{fig:maps_electrons} can be found in the Supplementary Materials.}.
The precipitation maps are obtained using $\sim 10^6$ macro-particles, enabling a good representation of the electron distribution function at the surface.

The electron precipitation maps in Fig.~\ref{fig:maps_electrons} show significant spatial inhomogeneities for both explored IMF configurations.
In the case of northward IMF (RunN), the solar-wind electrons are (i) energized up to energies of a few keV and (ii) concentrated onto the northern and southern cusps. In RunN, the northern cusp extends down to latitude $\sim+60^{\circ}$ while the southern cusp goes up to $\sim-30^{\circ}$. Energy fluxes in the cusps reach values of $\sim 10^{12}$ eV/cm$^{2}$s, two orders of magnitude higher than in the pristine solar-wind (Tab.~\ref{tab:tab1_pars}).
In the case of southward IMF (RunS), high-energy electrons (up to few keV) precipitate at low latitudes close to the magnetic equator (from -50$^{\circ}$ to +60$^{\circ}$) and mainly at the nightside.
This energy flux is higher in the dawn side (LT 0-6 h) as compared to the dusk side (LT 18-24 h). Such dawn-dusk asymmetry is a consequence of the dawnward drift of electrons injected from the neutral line in the tail towards the planet~\citep{Christon1987,Dong2019,Lavorenti2022a}. Such dawnward enhancements can also be due to a dawn-dusk asymmetry in the plasma sheet thickness~\citep[Chap. 4 therein]{Sun2022_review}.
In RunS, low-energy electrons (around tens of eV) precipitate around the northern and southern poles. These electrons precipitate directly from the solar wind onto the surface without crossing the reconnection region.

We find that 1.5 times more electrons precipitate on the surface in RunS as compared to RunN. 
The rate of precipitating electrons is $1.7\cdot 10^{25}$ s$^{-1}$ ($2.6\cdot 10^{25}$ s$^{-1}$) in RunN (RunS), corresponding to an effective area of 2\% (3\%) of the total planet surface area.
The rates, fluxes and energies reported here are in agreement with the findings of~\citet{Schriver2011b}.
For unmagnetized bodies (such as Mars, Venus, comets or the Moon), the effective area is 50\% of the planet surface area. In those cases, precipitation is much higher as compared to Mercury, but the solar wind does not suffer acceleration in the magnetosphere~\citep{Kallio2008}.
A weak magnetic field, like the one of Mercury, thus (i) filters the solar-wind in precise regions of the surface and (ii) accelerates the incoming particles by around a factor 100 in energy. Both effects (filtering and acceleration) are not possible around unmagnetized objects.

\begin{figure*}
    \centering
    \includegraphics[width=0.3\linewidth]{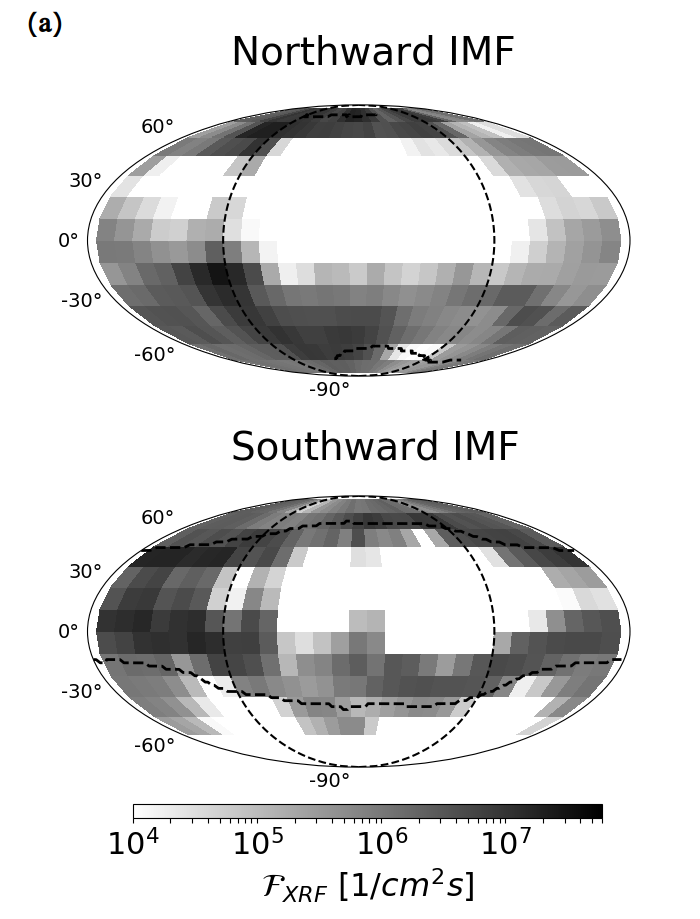}
    \includegraphics[width=0.5\linewidth]{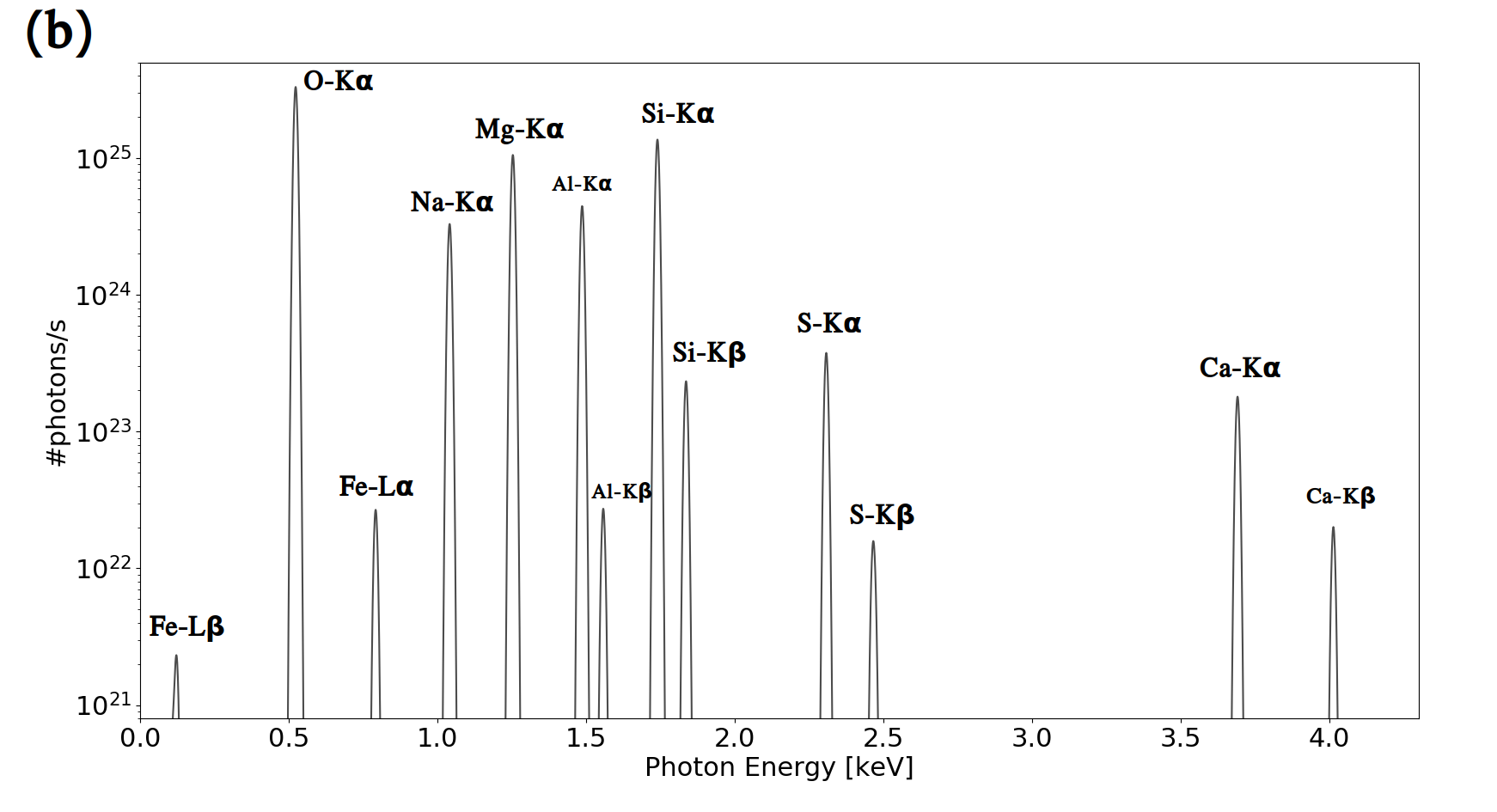}
    \caption{X-ray emission from the surface of Mercury due to electron precipitation. Panel a: maps showing the X-ray photon flux emitted from the surface of Mercury in our two runs. Maps in same format as Fig.~\ref{fig:maps_electrons}-\ref{fig:EII_maps}a.
    Panel b: X-ray photon flux integrated over surface area as a function of photon energy. Contributions from different surface atoms are shown. To better visualize the lines, we use a gaussian line profile with width $\sigma=5$ eV.}
    \label{fig:XRF_maps}
\end{figure*}

Before hitting the rocky surface of the planet, electrons interact with the exosphere. We address the efficiency of electron impact ionization (hereafter EII) of multiple exosphere species, using the electron energy distribution from our simulations, because we want to quantitatively estimate the efficiency of EII (usually considered a secondary, negligible effect) in comparison to photoionization (that is the primary ionization process).
For details on the method used, see the Appendix.
The distribution of EII rates over the planet's surface is shown in Fig.~\ref{fig:EII_maps}a for hydrogen (H), helium (He), oxygen (O) and manganese (Mn) for both simulation runs~\footnote{A more complete set of maps for all species considered can be found in the Supplementary Materials.}.
These four species have the highest EII-to-photoionization frequency ratio, as shown in Fig.~\ref{fig:EII_maps}b (photoionization rates $\nu_{ph}$ are taken from~\citet{Huebner2015} for quiet sun conditions and rescaled to Mercury's aphelion). For H, He, O and Mn, EII is relevant (i) in the dayside, where locally $\nu_{EII} \approx \nu_{ph}$ and (ii) in the nightside, where ionization of neutrals is dominated by EII with typical rates of $\sim$ 0.1 $\nu_{ph}$. The main difference between photoionization and EII remains the strongly inhomogeneous and time-dependent character of the latter. Variations in the IMF direction, such as moving from northward to southward IMF, induce strong variations in EII rates locally at the surface. 
We also show that EII of sodium (Na), magnesium (Mg), aluminum (Al), silicon (Si), potassium (K) and calcium (Ca) is negligible, as shown in Fig.~\ref{fig:EII_maps}b. This result support the common assumption of negligible EII for the Na exosphere~\citep{Sun2022,2021GeoRL..4892980J}. Nevertheless, compared to photoionization, EII should not be considered a secondary process for the H, He, O and Mn exosphere.

When hitting the surface, electrons induce the emission of photons from the surface atoms via X-ray fluorescence (hereafter XRF). XRF is driven by electrons with energies above a few hundreds eV~\citep[Tab. 5 therein]{Bunce2020}.
From the electron energy distribution at the surface obtained from our simulations, we compute the flux of emitted X-rays, as shown in Fig.~\ref{fig:XRF_maps}a. We include XRF emission from surface O, Na, Mg, Al, Si, S, Ca and Fe. In Fig.~\ref{fig:XRF_maps}a the photon flux is integrated over these species. Fig.~\ref{fig:XRF_maps}b, instead, shows the relative emission from each surface atoms as a function of photon energy, integrated over the planet surface.
The spatial distribution of surface atoms is assumed uniform, with density $n_s$.
The details on the method is described in the Appendix.
We find that auroral X-ray emissions from the surface of Mercury present strong spatial inhomogeneities dependent upon the upstream IMF conditions. Regions of strongest X-ray emission correspond to regions of high-energy electron precipitation, namely, the poles (in the case of northward IMF) and the low latitude dawn-midnight sector (in the case of southward IMF).
In these regions, the emitted X-ray flux reaches values of the order of 10$^7$ photons/cm$^2$s (mostly coming from the O-K$\alpha$ line).
MESSENGER/XRS instrument was able to measure X-rays from Si-K$\alpha$ and Ca-K$\alpha$ lines at Mercury's nightside~\citep{Lindsay2016}, and partially at the dayside~\citep{Lindsay2022} during periods of low solar activity. Our results corroborate the idea that Mercury's X-ray auroras are due to high-energy electron precipitation at the surface.
Moreover, our modeled X-ray flux provides a new mean to interpret and plan future in-situ observations by the BepiColombo/MIXS instrument~\citep{Bunce2020}.
The integrated X-ray luminosity of Mercury from electron-induced XRF is $\mathcal{L} \approx 10^{26}$ photons/s. Such luminosity is comparable (or even higher) to that of other solar system bodies shining in X-rays~\citep{Bhardwaj2007}, such as comets, Jupiter and Saturn. Nonetheless, remote observations from Earth of Mercury's X-ray aurora remains challenging due to the strong background of X-ray photons coming from the Sun.

To conclude, using a novel 3D fully-kinetic approach, we investigated the properties of solar-wind electron precipitation on the surface of Mercury-like bodies.
The magnetosphere of those bodies acts as (i) a shield, allowing only few percents of the solar wind to precipitate onto the surface, and (ii) an accelerator, increasing electron energies by a factor $\sim$100.
Using such precipitation pattern as the input to exosphere and surface impact processes, we find two main results.
First, electron impact ionization of exospheric H, He, O and Mn is efficient, with rates comparable to photoionization. This is particularly important on the nightside where photoionization is inhibited. The rates provided here are crucial to complement fluid and hybrid models of Mercury's environment, that cannot model electron acceleration processes.
Second, electron accelerated in the magnetosphere induce X-ray emission from the surface with fluxes of the order of $10^7$ photons/cm$^2$/s. This result corroborate and provide the physical origin of the X-ray auroras observations by MESSENGER/XRS instrument; it also paves the way for the future planning of BepiColombo/MIXS observations. 


\begin{acknowledgments}
This work was granted access to HPC resources at TGCC under the allocation A0100412428 and A0120412428 made by GENCI via the DARI procedure.
We acknowledge the CINECA award under the ISCRA initiative, for the availability of high performance computing resources and support for the project IsC93.
We acknowledge the support of CNES for the BepiColombo mission.
Part of this work was inspired by discussions within International Team 525: ``Modelling Mercury's Dynamic Magnetosphere in Anticipation of BepiColombo'' at the International Space Science Institute, Bern, Switzerland.
We acknowledge support by ESA within the PhD project ``Global modelling of Mercury's outer environment to prepare BepiColombo''. 
JD gratefully acknowledges support from NASA's Solar System Exploration Research Virtual Institute (SSERVI): Institute for Modeling Plasmas, Atmosphere, and Cosmic Dust (IMPACT), the NASA High-End Computing (HEC) Program through the NASA Advanced Supercomputing (NAS) Division at Ames Research Center, and NASA’s Rosetta Data Analysis Program, Grant No. 80NSSC19K1305.
\end{acknowledgments}

\appendix*

\section{Methods to compute EII rates and XRF emissions}
To compute electron impact ionization (EII) rates and X-ray fluorescence (XRF) emissions, we employ the electron energy distribution at the surface computed from our fully-kinetic simulations. 
We also use the cross sections from~\citet{NIST2005,Zatsarinny2019,Golyatina2021} for EII, and from~\citet{NIST2014} for XRF.
For a given process (EII or XRF with a given atomic species), the rate of interaction is:
\begin{equation}\label{eq:eq1_nuX}
    \nu_X (\phi,\theta) = \int^{\infty}_0 f_e (\phi,\theta,E) \sigma_X(E) \sqrt{\frac{2E}{m_e}}  dE
\end{equation}

where $f_e$ is the distribution of electrons at the surface as a function of longitude $\phi$, latitude $\theta$ and energy $E$ and $\sigma_X$ is the cross section of the process under consideration. Eq.~\ref{eq:eq1_nuX} is used to compute the EII rates in Fig.~\ref{fig:EII_maps}. We use the rate $\nu_{EII}$ to estimate the efficiency of EII with a given exosphere species because (i) it is independent on the density of target atoms and (ii) it is directly comparable to the photoionization rate $\nu_{ph}$, a value commonly used and well-known for multiple atomic species, e.g.~\citet{Huebner2015}.

To compute the X-ray flux emitted via particle-induced XRF from the surface, information on surface density and composition are needed. We employ the Mercury's geochemical surface composition of~\citet[Tab. 7.1]{McCoy2018} derived from MESSENGER observations. Using a mean surface mass density $n_{mean} = 3~\text{g/cm}^3$, we obtain the following densities for Si ($5~10^{22} \text{cm}^{-3}$), O ($4.5~10^{22} \text{cm}^{-3}$), Mg ($2.5~10^{22} \text{cm}^{-3}$), Al ($1.3~10^{22} \text{cm}^{-3}$), Na ($6.5~10^{21} \text{cm}^{-3}$), Ca ($4.6~10^{21} \text{cm}^{-3}$), S ($2.3~10^{21} \text{cm}^{-3}$) and Fe ($5~10^{20} \text{cm}^{-3}$).
We also assume an electron penetration depth of $\delta$ = 1~$\mu$m. 
The total X-ray photon flux is:
\begin{equation}\label{eq:eq2_flux-XRF}
    \mathcal{F}_{XRF} (\phi,\theta) = \delta \sum_s \nu_{_{XRF},s} (\phi,\theta) n_s 
\end{equation}

where $n_s$ is the density of a given surface species. 
The species-dependent, spatially integrated X-ray photon flux is:
\begin{equation}\label{eq:eq3_integral-flux-XRF}
    F_{XRF,s} = \delta \iint \nu_{_{XRF},s} (\phi,\theta) n_s \sin(\theta) d\theta d\phi 
\end{equation}

Eq.~\ref{eq:eq2_flux-XRF} is used to compute the flux in Fig.~\ref{fig:XRF_maps}a.
Eq.~\ref{eq:eq3_integral-flux-XRF} is used to compute the flux in Fig.~\ref{fig:XRF_maps}b.

\bibliography{biblio}

\end{document}